\documentclass[english,aps,prc,superscriptaddress,amsmath,amssymb,showpacs]{revtex4}
\usepackage[T1]{fontenc}
\usepackage[latin9]{inputenc}
\usepackage{bm}
\usepackage{amstext}

\makeatletter

\providecommand{\tabularnewline}{\\}

\@ifundefined{textcolor}{}
{%
 \definecolor{BLACK}{gray}{0}
 \definecolor{WHITE}{gray}{1}
 \definecolor{RED}{rgb}{1,0,0}
 \definecolor{GREEN}{rgb}{0,1,0}
 \definecolor{BLUE}{rgb}{0,0,1}
 \definecolor{CYAN}{cmyk}{1,0,0,0}
 \definecolor{MAGENTA}{cmyk}{0,1,0,0}
 \definecolor{YELLOW}{cmyk}{0,0,1,0}
 }

\makeatother

\usepackage{babel}
\begin{document}

\title{Isobaric analog state in $\bm{^{96}}$Ag}

\author{L. Zamick}

\author{A. Escuderos}

\affiliation{Department of Physics and Astronomy, Rutgers University, Piscataway,
New Jersey, 08854}

\author{S.J.Q. Robinson}

\affiliation{Department of Physics, Millsaps College, Jackson, Mississippi, 39210}

\author{Y. Y. Sharon}

\affiliation{Department of Physics and Astronomy, Rutgers University, Piscataway,
New Jersey, 08854}
\begin{abstract}
Previously, in a single-$j$-shell calculation ($j=g_{9/2}$), we
obtained the excitation energy of the $J=0^{+}$, $T=2$ isobaric
analog state in $^{96}$Ag to be a bit below 1~MeV relative to the
$J=8^{+}$, $T=1$ ground state. We here use binding energy data and
Coulomb energy estimates to obtain this same excitation energy and
to see if the two approaches are consistent. 
\end{abstract}
\maketitle

\section{Results}

If there were no violation of charge independence, the binding energy
of $^{96}$Pd ground state ($J=0^{+}$, $T=2$) would be identical
to the binding energy of the analog state, also $J=0^{+}$, $T=2$,
in $^{96}$Ag. But, since that is not the case in real life, the excitation
energy of the $J=0^{+}$, $T=2$ state in $^{96}$Ag is then given
by 
\begin{equation}
E^{*}(J=0^{+},T=2)=BE(^{96}\text{Ag})-BE(^{96}\text{Pd})+V_{C}\,,\label{eq:exc}
\end{equation}
 where the $BE$s are the binding energies and $V_{C}$ includes all
charge-independence violating effects. We here assume that $V_{C}$
arises from the Coulomb interaction and use the formula of Anderson
et al.~\cite{awm65}: 
\begin{equation}
V_{C}=E_{1}Z/A^{(1/3)}+E_{2}\,,\label{eq:vc}
\end{equation}
 where $Z=(Z_{1}+Z_{2})/2$. Anderson et al.~\cite{awm65} list four
sets of values of $E_{1}$ and $E_{2}$. We here use the average values
$E_{1}=1.441$~MeV and $E_{2}=-1.06$~MeV.

We show in Table~\ref{tab:exc} results for various nuclei, some
for which the excitation energy of the analog state is known and some
for which it is not. The binding energy differences are taken from
Ref.~\cite{awt03}

\begin{table}
\caption{\label{tab:exc}Excitation energies of isobaric analog states in MeV.}

\begin{ruledtabular} %
\begin{tabular}{ccccccc}
NUCLEUS  & Binding Energy Difference  & Coulomb Energy  & Excitation Energy  & Single $j$  & Large space  & Experiment \tabularnewline
\hline 
$^{44}$Sc  & 4.435  & 7.308  & 2.873  & 3.047\footnotemark[1]  & 3.418\footnotemark[2]  & 2.779 \tabularnewline
$^{46}$Sc  & 2.160  & 7.184  & 5.024  & 4.949\footnotemark[1]  & 5.250\footnotemark[2]  & 5.022 \tabularnewline
$^{52}$Mn  & 5.494  & 8.399  & 2.905  & 2.774  & 2.7307  & 2.926 \tabularnewline
$^{60}$Cu  & 6.910  & 9.430  & 2.520  & 2.235 &  & 2.536 \tabularnewline
$^{94}$Rh  & 10.386  & 13.043  & 2.657  & 1.990\footnotemark[3]  & 3.2664\footnotemark[4] 2.87943\footnotemark[6]  & \tabularnewline
 &  &  &  & 2.048\footnotemark[3]  &  & \tabularnewline
$^{96}$Ag  & 12.432  & 13.574  & 1.142  & 0.900\footnotemark[3]  & 1.91667\footnotemark[4] 1.64017\footnotemark[6]  & \tabularnewline
 &  &  &  & 0.842\footnotemark[5]  &  & \tabularnewline
\end{tabular}\end{ruledtabular} \footnotetext[1]{Escuderos, Zamick, Bayman (2005)~\cite{ezb05}.}
\footnotetext[2]{GXPF1 interaction~\cite{hobm04}.} \footnotetext[3]{Zamick
and Escuderos (2012)~\cite{ze12}.} \footnotetext[4]{jj44b interaction~\cite{bl-un}.}
\footnotetext[5]{CCGI interaction~\cite{ze12,ccgi12}.} \footnotetext[6]{}JUN45
interaction~\cite{jun45} 
\end{table}

The fact that the analog state and Coulomb arguments work well in
known cases gives us confidence that we can use these for the unknown
case of $^{96}$Ag. Turning things around, if the isobaric analog
state were found, then we might have a better constraint on what the
binding energy is.

We can compare the results of the calculated excitation energies with
selected calculations in the literature. For $^{44}$Sc and $^{46}$Sc,
single-$j$-shell results ($f_{7/2}$)~\cite{ezb05} are respectively
3.047 and 4.949~MeV, as compared with Table~\ref{tab:exc}'s results
of 2.873 and 5.024~MeV. The large space results are also shown. In$^{52}$Mn
there is resonable agreement beweencalculated, single j, large space
and experiment.

For the small space for $^{60}$Cu (p$_{3/2}$) we can use a particle
-hole transformation to get the specrum of this nucleus from the spectrum
of $^{58}$Cu since 3 p$_{3/2}$ neutrons can be regarded as a single
neutron hole. Ths gives a value of 2.235 MeV as compares with experiment--2.536
MeV.

For $^{96}$Ag single-$j$-shell results~\cite{ze12} are 0.900~MeV
with INTd and 0.842~MeV with the CCGI interaction~\cite{ze12,ccgi12}.
These are lower than the value in Table~\ref{tab:exc} of 1.142~MeV.
There are also large scale calculations with the jj44b~\cite{bl-un}
interaction for $^{96}$Ag---the result is 1.996~MeV, significantly
larger than the calculated value. In $^{94}$Rh the jj44b interaction
yields 3.052~MeV, larger than the Table~\ref{tab:exc}'s value of
2.657~MeV. The large space calculations with June45 are qualitatively
similar.The single- j INTd and CCGI results are lower, 1.990 MeV and
2.048 MeV respectively.

We can also examine this problem usng various mass formulas that abound
in the literature. To this end we refer to the work of Kirson{[}17{]}
which contains not only the parameters of the semiemprical mass formula
of Bethe and Weisacker {[}18{]} but also a more elaborate formula
that he develped. Also to be considered is the mass formula of Dulfo
and Zuker {[}19{]} which is generaaly considerd to be the best on
the market.

We here present the results of the excitation energies in the format
Nucleus (semiempirical, Zuker, KirsonA, KirsonB, repeat of table 1).
In semiempirical and KirsonA we use the Coulomb energies contained
the respective mass formulas. In Zuker and KirsonB we use the Coulomb
energies from Table 1 {[}1{]}.

$^{44}$Sc (2.526, 2.374, 1.947, 2.592 , 2.873)

$^{46}$Sc (6.250, 4.744, 4.532, 5.060, 5.024 )

$^{52}$Mn(1.875, 1.927, 1.418, 1.911, 2.905 )

$^{60}$Cu(1.408, 2.420, 1.013, 1.514, 2.520 )

$^{94}$Rh(2.316, 2.205, 1.503, 1.734, 2.657)

$^{96}$Ag(0368, 0.689, -0.036, 0.173,1.142 )

We next list the Coulomb energy difference for (seiemprircal, Kirson,
Table )

$^{44}$Sc (8.025, 6.663, 7.308)

$^{46}$Sc (7.906,6.652,7.184)

$^{52}$Mn (9.071,7.827,8.399)

$^{60}$ Cu(10.061, 8.928, 9.430 )

$^{94}$Rh (13.526, 12.812, 13.043)

$^{96}$Ag (14.035,13.364,13.547)

The Kirson value is smaller than the semiempirical one because it
includes an exchange term. In the furture it would be useful to get
a better handle on the Coulob energies.

In view of the differing results of shell model calculations and mass
formulas it would be of great interest to measure the excitation energies
of isobaric analog states in the $g_{9/2}$ region. We hope that this
work will encourage experimentalists to look not ony for the surprisingly
neglected $J=0^{+}$ isobaric analog states in $^{94}$Rh and $^{96}$Ag,
but also for other such states throughout this region. 
\begin{acknowledgments}
We thank Klaus Blaum for his help. We are indebted to Michael Kirson
for valuable discussions and for providing us with the Dulfo- Zuker's
results.\end{acknowledgments}

\end{document}